\documentclass[12pt]{iopart}

\usepackage[dvips]{graphicx}
\usepackage{iopams}

\def\ket#1{|#1\rangle}
\def\bra#1{\langle#1|}
\def\ketbra#1{|#1\rangle\langle#1|}
\def\idmat{\mathbf{1}}
\def\caln{\mathcal{N}}
\def\calc{\mathcal{C}}

\def\tr{\mathrm{tr}}
\def\bfu{\mathbf{u}}
\def\bfn{\mathbf{n}}
\def\bfni{\mathbf{n_i}}
\def\bfnj{\mathbf{n_j}}

\def\dm{\frac{1}{2}}
\def\nae{\textrm{n.a.e.}}

\newcommand{\bJ}{{\bf J}}

\newcommand{\cC}{{\cal C}}
\newcommand{\cN}{{\cal N}}
\newcommand{\ovtheta}{{\bar{\theta}}}

\begin{document}

\title{Quantifying Quantumness and the Quest for Queens of Quantum}
\author{Olivier Giraud$^{1,2,3,4}$, Petr A.~Braun$^{5,6}$ and Daniel Braun$^{1,2}$}
\address{$^{1}$ Universit\'e de Toulouse; UPS; Laboratoire de
 Physique Th\'eorique (IRSAMC); F-31062 Toulouse, France\\
$^{2}$ CNRS; LPT (IRSAMC); F-31062 Toulouse, France\\
$^{3}$ Universit\'e Paris-Sud, LPTMS, UMR8626, B\^at. 100, Universit\'e Paris-Sud, 91405 Orsay, France\\
$^{4}$ CNRS,  LPTMS, UMR8626, B\^at. 100, Universit\'e Paris-Sud,
  91405 Orsay, France\\
$^{5}$ Fachbereich Physik, Universit\"at Duisburg--Essen, 47048 Duisburg,
  Germany\\
  $^{6}$ Institute of Physics, Saint-Petersburg University,
198504 Saint-Petersburg,  Russia}
\begin{abstract}We introduce a measure of ``quantumness'' for any quantum
  state in a finite dimensional Hilbert space, based on the distance
  between the state and the convex set of
  classical states. The latter are defined as states that can
  be written as a convex sum of projectors onto coherent states.  We derive
  general properties of this measure of non-classicality, and use it to
  identify for a given
  dimension of Hilbert space what are the ``Queen of Quantum'' states,
  i.e.~the most 
  non-classical quantum states.  In three dimensions we obtain the Queen
  of Quantum state analytically and show that it is unique up to rotations.
  In up to 11-dimensional Hilbert spaces, we find the Queen of Quantum 
  states numerically, and show that in terms of their Majorana
  representation they are highly symmetric bodies, which for
  dimensions 5 and 7 correspond to Platonic bodies.
\end{abstract}
\maketitle

\section{Introduction}
The advent of quantum information theory has led to substantial efforts to
understand the resources which are responsible for the enhanced information
processing capabilities of quantum systems compared to classical ones.  A
large part of that research has been directed towards the creation and
classification of entanglement
\cite{Lewenstein00,Plenio05}. Entanglement
plays an important role in quantum teleportation \cite{Bennett93} and various
quantum communication schemes \cite{Nielsen00}.  It is also known that any
pure state
quantum computation which does not produce large scale entanglement can be
simulated efficiently classically \cite{Jozsa03}.  Physically, entanglement
manifests itself as increased correlations between different subsystems
compared to what is possible classically \cite{Bell64,Gisin91}.  But even
for a system consisting of only a single subsystem
one may ask how ``quantum'' a given state is, and what benefits one might
draw from its ``quantumness''.

In physics there is a wide
consensus that the ``least quantum'' (or ``most classical'') pure states are
coherent states. These are states which present the
smallest possible amount of quantum fluctuations, as defined by a suitable
Heisenberg uncertainty relation, evenly distributed over a
pair of non-commuting variables. For example, in quantum optics, coherent
states have
that property of minimal and equal uncertainty for the field quadratures.
Moreover, the 
dynamics of the latter is identical to that given by the classical equations
of motion of the harmonic oscillator, and the property of minimal
uncertainty is conserved during the time evolution created by the
Hamiltonian of the electromagnetic field. The most classical mixed states
possible can be obtained as a statistical mixture of coherent states. Any
mixed state can be expanded over projectors on coherent states, with real
coefficients given by the so-called $P$-function. 
Consequently, in quantum optics, states
with a positive $P$-function are considered as ``most classical''~\cite{Kim05,Mandel86}.

In
\cite{Giraud08} we extended that definition to systems with
Hilbert spaces of finite dimension $d$. These systems are formally
equivalent to a spin of size $j$ with
$d=2j+1$. The $P$-function is not uniquely determined in that case, but
leaves a lot of freedom in the specification of the higher spherical harmonics
components. We therefore defined the set ${\calc}$ of
classical states (also called the set of ``P-rep'' states in \cite{Giraud08},
for positive
$P$-function) as the ensemble of all density matrices $\rho_c$ for which a
decomposition in terms of angular momentum coherent states
$|\ovtheta\varphi\rangle$ 
(see (\ref{zetaC}) for a precise definition) with positive weights exists,
\begin{equation} \label{prep}
\rho_c=\sum_i \mu_i|\ovtheta_i\varphi_i\rangle\langle\ovtheta_i\varphi_i| 
\end{equation}
with $0\le \mu_i\le 1$ and $\sum_i \mu_i=1$.  At most $d^2=(2j+1)^2$ terms
 in the convex sum are needed \cite{Giraud08}. We showed that for $j=1/2$
 all states are classical in that sense.  For spin $j=1$ we found a
 necessary and sufficient criterion for 
 classicality, and for higher values of $j$ we found ``non-classicality
 witnesses'' which allow to easily detect large classes of non-classical states
 through the violation of necessary conditions for classicality derived
 from (\ref{prep}).
For composite systems, the set of classical states $\calc$ is in general
 strictly smaller than the set of separable states \cite{Giraud08}.

The definition (\ref{prep}) of a classical state allows so far just to
determine whether a state is classical or not.  However, it would be
interesting to know {\em how} ``non-classical'' (or, in other words, how
``quantum'') a given state is, as one might expect that very non-classical
states might be more useful for applications in quantum information
processing than states which are only slightly non-classical.    The
situation is
very analogous to the one encountered in the study of entanglement, where
one wants to have an entanglement measure in addition to
entanglement criteria.

This is the question we are going to pursue in this work.  We introduce a
measure of quantumness in the next section, study some of its
properties, and then apply it to find the ``most quantum''
states possible for a given Hilbert space dimension. We show that the
largest possible quantumness can always be found in a
pure state. The states with maximal quantumness turn out
to possess remarkable geometrical beauty.  We term them 
``Queens of Quantum'' (QQ) states.  In
the lowest-dimensional non-trivial case ($j=1$, i.e.~$d=3$) there is a
unique Queen of Quantum (up to rotations of the coordinate system),
which we determine analytically.  In higher dimensions (up to $j=5$) we find
the Queens of Quantum numerically using quadratic optimization.
Other attempts to define the ``least classical'' quantum states were
proposed in the literature based on properties of the average value
and the variance of
the (pseudo-)angular momentum operator $\bJ$
\cite{Davis00,Zimba06,Kolenderski08}. We will 
briefly discuss these results in relation to the QQ states.  

\section{Measure of Quantumness}
\subsection{Definition and properties}\label{sec.defprop}
We define the ``quantumness'' $Q(\rho)$ of an
arbitrary state $\rho$ as the distance from $\rho$ to the convex set of
classical states
$\calc$.
We thus introduce the measure of quantumness by defining
\begin{equation}
Q(\rho)\equiv \min_{\rho_c\in \calc}||\rho-\rho_c||\,,\label{defNC}
\end{equation}
where the minimum is over all classical states defined in (\ref{prep}), and
$||A||\equiv \tr(A^\dagger A)^{1/2}$ denotes the Hilbert-Schmidt norm. Note
that our 
definition of quantumness is very analogous to the entanglement measure
based on the
distance of a state $\rho$ from the convex set of separable states
\cite{Vedral98}.  

Several consequences follow immediately from (\ref{defNC}):\\

1. For any state $\rho$, and any dimension $d$, we have the bounds
  $0\le Q(\rho)\leq \sqrt{\tr\rho^2}+\sqrt{\tr\rho_c^2}\leq 2$. The lower
  bound   is trivially realized for classical
  states. This implies $Q(\rho)=0$ for all states of a spin 1/2.\\

2. An improved upper bound on $Q(\rho)$ that only depends on the purity
  of $\rho$ can be found by
  considering the
  distance to the maximally mixed state ${\bf 1}/(2j+1)$, which is always
  classical \cite{Giraud08}. This immediately leads to
\begin{equation} \label{ub2}
Q(\rho)\le \sqrt{\tr\rho^2-\frac{1}{2j+1}}\,.
\end{equation}
For a pure state this bound coincides with the less stringent 
\begin{equation}
\label{upper_bound_pure}
Q(\rho)\le\sqrt{1-\frac{1}{2j+1}}.
\end{equation}
3. A different upper bound can be found by minimizing over a
  single pure coherent state $|\alpha\rangle=\ket{\ovtheta\varphi}$:
\begin{equation} \label{hub}
Q(\rho)\le
\min_{\alpha}\vert\vert\rho-|\alpha\rangle\langle\alpha|\,\vert\vert\le
(1+\tr\rho^2-2\max_{\alpha}H_\rho(\alpha))^{1/2}\,,
\end{equation}
where $H_\rho(\alpha)\equiv\bra{\alpha}\rho\ket{\alpha}$ denotes the Husimi
function of the state. For a pure state
the bound becomes $Q(|\psi\rangle\langle\psi|)\le
\sqrt{2}(1-\max_{\alpha}|\langle\alpha|\psi\rangle|^2)^{1/2}$.\\

4. As the distance to a convex set is a convex function
 (see e.g.~example 3.16 of~\cite{Boyd04}), $Q(\rho)$ is a convex function, i.e.~for any two
  states $\rho_1$, $\rho_2$, and $0\le p\le 1$, we have
\begin{equation}
Q(p\rho_1+(1-p)\rho_2)\le p Q(\rho_1)+(1-p)Q(\rho_2)\,.\label{convQ}
\end{equation}
This implies that
  the quantumness of any mixed state cannot be larger than the largest
  quantumness of the pure states of which it is a mixture.\\

5. $Q(\rho)$ is invariant under rotations of the coordinate
  system.  Indeed, let $R_\bfn=\exp(i \bfn.\bJ)$ be a unitary operator
  associated with a rotation of the coordinate system about the axis
  $\bfn$ by an angle $|\bfn|$. Since $R_\bfn$ is unitary, we have
  $||\rho-\rho_c||=||R_\bfn\rho R_\bfn^\dagger-R_\bfn\rho_c R_\bfn^\dagger||$
  for all
 density matrices $\rho,\rho_c$.  Furthermore, for $\rho_c\in \cC$, we
  also have $\tilde{\rho}_c\equiv R_\bfn\rho_c R_\bfn^\dagger\in
  \cC$, as $R_\bfn$ only rotates coherent states into other coherent
  states, and therefore does not change the classicality of $\rho_c$.
  Moreover, the map $\rho_c\to\tilde{\rho}_c$ for given $R_\bfn$ is an
  isomorphism 
  $\cC\to\cC$. Therefore, we have $\min_{\rho_c\in
  \cC}=\min_{\tilde{\rho}_c\in\cC}$. This leads to
  $Q(\rho)=\min_{\rho_c\in\cC}||\rho-\rho_c||=
  \min_{\tilde{\rho}_c\in\cC}||R_\bfn\rho
  R_\bfn^\dagger -\tilde{\rho}_c||=Q(\tilde{\rho})$ for $\tilde{\rho}=R_\bfn\rho
  R_\bfn^\dagger$.\\

With the same argument one shows that for a composite system consisting of
  $s$ subsystems, $Q(\rho)$ is
  invariant under independent rotations for all the subsystems, i.e. under
  transformations $R=R_{\bfn_1}\otimes R_{\bfn_2}\otimes \ldots
  R_{\bfn_s}$. In addition, for a 
  system consisting of $s$ qubits, $Q(\rho)$ is invariant under all local
  unitary transformations, as in this case
  local unitary $SU(2)$ transformations leave the set $\calc$
  invariant.  Our measure of classicality shares this 
  property with any measure of entanglement. Indeed, for a multi
  spin-1/2 system, the set of classical states $\cC$ is identical to
  the set of totally separable states. Therefore, in this case
  $Q(\rho)$ coincides with an entanglement measure.
For higher-dimensional
  subsystems, this is, of course, not true, as for example a $SU(3)$
  transformation of a single spin 1 can transform a coherent state into a
  non-classical state.

\subsection{Simple examples}
\subsubsection{Spin-1/2 case}
As mentioned in \cite{Giraud08}, any spin-$\frac{1}{2}$ pure state is
classical, thus the set $\calc$ coincides with the set of all quantum
states. All states are thus trivially at distance 0 from $\calc$.

\subsubsection{Pure spin-1 case}
\begin{figure}
\begin{center}
\vspace{1cm}
\includegraphics[scale=.35]{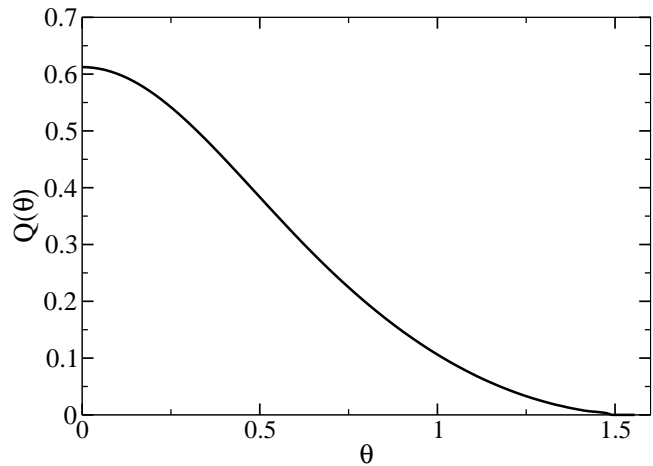}
\end{center}
\caption{Quantumness of the state (\ref{canonicalj1}) as function of $x=\sqrt{2}/\sin\theta$. The plot of (\ref{distancex}) coincides with this curve within numerical accuracy. \label{figure0}}    
\end{figure}

In order to illustrate the behaviour of the quantumness in the simplest
non-trivial case, let us consider the family of pure spin-1 states given by
\begin{equation}
\label{canonicalj1}
\ket{\psi_x}=\frac{1}{\sqrt{x^2+2}}
\left(\ket{1,-1}+x\ket{1,0}+\ket{1,1}\right)\,,
\end{equation}
where $\ket{jm}$, $-j\leq m\leq j$, are eigenvectors of
the spin angular momentum operator $J_z$ with eigenvalue $m$.
Let $x=\sqrt{2}/\sin\ovtheta$. In figure \ref{figure0} we plot the quantumness $Q(\ket{\psi_x}\bra{\psi_x})$
obtained numerically  
using the method described in section \ref{numerics}.
We see that the largest quantumness is obtained at $\ovtheta=0$ (or $x=\infty$), corresponding
to the state $\ket{1,0}$. We  will
prove in section \ref{spin1} that this state is indeed the spin-1 Queen of
Quantum state and has $Q=\sqrt{3/8}$.

\subsubsection{Thermal spin states}
Consider a density matrix given by the thermal state $\rho=\exp(-\beta
H)/\tr(\exp(-\beta H))$ with inverse temperature $\beta=1/k_BT$ and Hamiltonian
$H$. For $\beta\to 0$ all energy eigenstates are equally populated, so
that $\rho$ is the (properly normalized) identity matrix $\rho_0$
corresponding to the maximally 
mixed state. Since $Q(\rho_0)=0$, we have the intuitively appealing result
that for sufficiently large 
temperature all thermal states become classical. Indeed, there is a finite
temperature where this happens, as there is a finite neighbourhood around
$\rho_0$ (with a finite radius in all directions) where all states are classical \cite{Giraud08}.  For  $\beta\to
\infty$ the quantumness depends on the quantumness of the ground state(s) of
$H$.  

We will illustrate what can happen with two examples for spin 1. In this
case we have at 
our disposal a necessary and sufficient condition of classicality, which has
been obtained in  
\cite{Giraud08} (see Sect.~\ref{spin1} for a presentation of this criterion). \\

1. For $H=J_z$ the ground state is the pure coherent state $\ket{1,-1}$ and
thus it is classical. Our classicality criterion shows that
in this case all thermal states are actually classical.\\

2. For $H=J_z^2$, the ground state is the non-classical state $\ket{1,0}$,
and the classicality criterion shows that there is a critical temperature
$T_c=1/\beta_c=1/\ln 2$ above which the quantumness disappears. \\

Therefore, while the quantumness at low temperature depends on the model and
in particular on the quantumness of low-lying states, classicality is a
universal property for 
systems at thermal equilibrium for $T\to\infty$.


\section{Queens of Quantum}
We now address the question of finding out which states are the ``most
quantum'' states. This is a highly non-trivial question, as it
requires to find a state that maximizes a quantity defined as a
minimum over a convex set. We first introduce some definitions.
\subsection{Definitions and general properties}
We define the ``Queen of Quantum'' (QQ) states as those states
$\rho_{QQ}\in \cN$ for
which $\rho_{QQ}=\max_{\rho\in\cN}Q(\rho)$, where $\cN$ is the set of
all physical density operators acting on a Hilbert space of given
finite dimension.  In other words, the Queens of Quantum are the
``most quantum'' states for a given Hilbert space dimension.
\subsubsection{Pure versus mixed}
In order to identify the largest quantumness possible for a given
Hilbert space dimension, we can restrict ourselves
to pure states, as according to (\ref{convQ}) the quantumness of a
state $\rho$ cannot be larger than the quantumness of the pure states
from which it is a mixture. This does not immediately imply, however, that
all QQ states are pure.  Indeed, suppose that two pure states $|\psi_1\rangle$
and $|\psi_2\rangle$ have the same quantumness,
$Q(|\psi_1\rangle\langle\psi_1|)=Q(|\psi_2\rangle\langle\psi_2|)$.
Then, according to (\ref{convQ}),
$Q(p|\psi_1\rangle\langle\psi_1|+(1-p)|\psi_2\rangle\langle\psi_2|)\le
Q(|\psi_1\rangle\langle\psi_1|)$
for $0\le p\le 1$. Equality is possible in principle, unless $Q(\rho)$
is strictly convex, which, however, need not be the case.  For example
one might imagine that $\cN$ has a flat surface containing
$|\psi_1\rangle\langle\psi_1|$ and $|\psi_2\rangle\langle\psi_2|$,
in parallel to a flat surface of $\cC$ containing the two closest mixed
states for $|\psi_1\rangle\langle\psi_1|$ and
$|\psi_2\rangle\langle\psi_2|$.
In this case, all states mixed from
the two pure QQ states will have the same maximum quantumness.
The problem of degenerate
quantumness of the QQ states is in fact generic, as we have seen that
all states obtained by rotation of the coordinate system have the same quantumness.
Nevertheless, we can start off by determining all pure QQ states, and
then try to determine whether states can be mixed from these which
would give the same maximum quantumness.
For $j=1$ we will show explicitly that all QQ states are pure.

\subsubsection{Majorana representation}
We now restrict ourselves to the case where $\rho=\ket{\psi}\bra{\psi}$ is a
pure state. A spin-$j$ coherent state can be expressed as \cite{Agarwal81}
\begin{equation}
\ket{\ovtheta \varphi}=\sum_{m=-j}^{j}{2j \choose j+m}^{\dm}
\left( \sin \frac{\ovtheta }{2}\right) ^{j+m}\left( \cos \frac{\ovtheta
}{2}\right) ^{j-m}e^{-i(j+m)\varphi }\ket{jm}.
\label{zetaC}
\end{equation}
Note that with this definition the state $\ket{j\,-j}$ corresponds to
$\ovtheta=0$.  We take this state as the South pole, and $\ket{j\,j}$ as the
North pole.
The state $\ket{\ovtheta \varphi}$ can be seen as the state obtained by
rotating $\ket{j\,-j}$ in a direction specified on the sphere $S^{2}$ by
an angle $\ovtheta$ about the $y$-axis followed by an angle $\varphi$ about the
$z$-axis.  This means that the mean
value $\bra{\ovtheta \varphi}\bJ\ket{\ovtheta \varphi}$ of the angular momentum
vector points in the direction given by the usual (i.e. counted from
the North pole) polar angle $\theta=\pi-\ovtheta$ and
azimuth $\varphi$. We draw the attention of the reader to the fact
that in the whole paper $\ovtheta$ will be counted from the South
pole. This perhaps unusual convention is adopted here in order to
simplify subsequent expressions for the Majorana polynomials.

Any pure
state $\ket{\psi}$ can be represented by its overlap with coherent states
\begin{eqnarray} \label{majfun}
\langle\ovtheta\varphi|\psi\rangle&=&\frac{1}{(1+|\zeta|^2)^j}\sum_{m=-j}^{j}
{2j \choose j+m}^{\dm}\psi_{m}\zeta^{j+m} \,,
\end{eqnarray}
where $\zeta$ is a complex number defined by
\begin{equation}
\label{zeta}
\zeta=e^{i\varphi}\tan\frac{\ovtheta}{2}.
\end{equation}
The scalar product (\ref{majfun}), up to a prefactor
independent of $\ket{\psi}$, is a polynomial of degree at most
$2j$ in the complex variable $\zeta$. If $\psi_j\ne 0$, it is a polynomial
of degree $2j$, and has $2j$ complex roots $\zeta_1,\ldots\zeta_{2j}$. This
so-called Majorana polynomial \cite{Majorana1932, Penrose89} reads
\begin{equation} \label{Mz}
M(\zeta)=\sum_{k=0}^{2j}
  {2j \choose k}^{\dm}\psi_{k-j}\zeta^k=\psi_j\prod_{k=1}^{2j}(\zeta-\zeta_k)\,.
\end{equation}
The inverse stereographic projection of these $2j$ roots
define $2j$ points on the unit sphere $S^2$ through (\ref{zeta}). If the 
degree of $M(\zeta)$ is $D<2j$, the Majorana representation is defined as
consisting of $D$ 
points associated with the $D$ roots of the polynomial and $2j-D$ points at
the North pole, and the prefactor $\psi_j$ in (\ref{Mz}) is replaced by
$\psi_{D-j}$. 
This set of points is called
the Majorana  (or stellar) representation.
It entirely characterizes the
normalized state $\ket{\psi}$ up to a global phase.
A nice feature of this
representation is that all points rotate rigidly
if $\ket{\psi}$ undergoes a rotation
$R_\bfn=\exp(i \bfn.\bJ)$ of the reference frame.  If the state $\ket{\psi}$
is a coherent state, $\ket{\psi}=\ket{\ovtheta  \varphi}$, then the $2j$ roots
of 
its Majorana polynomial are all equal 
and given by 
$\zeta_k=\tan\frac{\pi-\ovtheta}{2}e^{i(\varphi+\pi)}$. They correspond to
$2j$ points  which are antipodal to
the vector $\bra{\ovtheta \varphi}\bJ\ket{\ovtheta \varphi}$.
For instance, the state
$\ket{j\,j}$ (corresponding to $\ovtheta=\pi$), which is a coherent
state whose mean value is located at the North pole, is represented by $2j$ points at
the South pole (once more we recall that we count
$\ovtheta$ from the South  pole).  \\

There is a natural interpretation of the Majorana representation in terms of
tensor products of spin states.
As a spin-$j$ state, $\ket{\psi}$ can be seen as a fully
symmetrized direct product of $2j$ spins $\frac{1}{2}$. Each
spin-$\frac{1}{2}$ state is a coherent state of the form (\ref{zetaC}). As
such it  can
be represented by a Majorana point on the Bloch sphere in the direction
antipodal to the coherent state.
The state $\ket{\psi}$ corresponds to a symmetrization of $2j$ Bloch spheres
and thus can be represented by a set of $2j$ points on a single sphere.
The Majorana representation is useful in many contexts:
calculation of the Berry phase for pure states \cite{Hannay98}, proof
of Sylvester's theorem on Maxwell multipoles \cite{Dennis04}, investigation of
states which maximize the variance $(\Delta{\bf J})^2=\langle {\bf J}.{\bf
  J}\rangle-\langle {\bf J}\rangle\langle{\bf J}\rangle$ \cite{Davis00},
or states such that this variance is uniform over the unit sphere
\cite{Zimba06}.\\

Since the set $\calc$ of classical states
is invariant under rotation (as coherent states are just
rotated to coherent states) the distance from $\ket{\psi}$ to $\calc$ is the
same for any rigid rotation of the $2j$ points on the sphere.
The problem of identifying pure QQ states for spin $j$ reduces to identifying optimal
distributions of $2j$ points on the unit sphere, up to global rotation.

\subsubsection{Pure state as eigenfunction of its classical neighbour}
\label{rhocisev}
Before proceeding to the identification of the most quantum states, we
derive an important property of QQ states.
As we pointed out, the largest $Q(\rho)$ can always be reached with a pure
state, $\rho 
=\left\vert \psi \right\rangle \left\langle \psi \right\vert $. The squared
distance from a pure state to $\mathcal{C}$ is maximized by the state
$\psi_{QQ}$ such that
\begin{equation}
\label{maximin}
Q^{2}\left(\ket{\psi_{QQ}}\bra{\psi_{QQ}}\right) =\max_{\ket{\psi}}\min_{\rho _{c}\in \mathcal{C}}\left(
1-2\left\langle \psi \left\vert \rho _{c}\right\vert \psi \right\rangle +
\mathrm{tr}\rho _{c}^{2}\right).
\end{equation}
The necessary condition on state $\psi_{QQ}$ is stationarity of
$Q^{2}\left(\ketbra{\psi}\right) $ with respect to variations of $\ket{\psi}$
constrained by the condition $\left\langle \psi |\psi
\right\rangle =1$.  The first variation of the functional
$I[\psi]=1-2\left\langle \psi |\rho _{c}|\psi \right\rangle+2E\left\langle \psi |\psi
\right\rangle $, equal to
\begin{equation}
\delta I=-2\langle
\delta\psi|\rho_c|\psi\rangle+2E\langle\delta\psi|\psi\rangle+\mathrm{c.c}\,,
\end{equation}
where $2E$ is the Lagrange multiplier,  must be zero for all
$\ket{\delta\psi}$. It follows that
\begin{equation}
\rho _{c}\ket{\psi} =E\ket{\psi} .
\end{equation}
Consequently, the wavefunction of a QQ state is an eigenfunction of (the
density matrix of) its nearest classical state. We checked that all QQ
states obtained in the following sections indeed do satisfy this
property.

\subsection{Spin-1 case}\label{spin1}
We now turn to the analytic investigation of the simplest nontrivial
case of spin 1.
\subsubsection{Pure states}
Let $J_{a}$, $a=x,y,z$, be the $3\times 3$ angular momentum matrices for
$j=1$. One can expand any density matrix $\rho$ in the basis of $J_{a}$ and
$(J_{a}J_{b}+J_{b}J_{a})/2$, as
\begin{equation}
\rho =\frac{1}{3}\idmat_{3}+\frac{1}{2}\mathbf{u}.\mathbf{J}
+\frac{1}{2}\sum_{a,b=x,y,z}\left( W_{ab}-\frac{1}{3}\delta
_{ab}\right) \frac{J_{a}J_{b}+J_{b}J_{a}}{2}\,.
\label{canonrhoj1}
\end{equation}
Here $\bfu$ is a real vector, $W$ a real symmetric matrix with trace 1, and
$\idmat_{3}$ the $3\times 3$ identity matrix.
They are related to $\rho$ by
\begin{equation}
\label{uW}
u_{a}=\tr\left( \rho J_{a}\right) ,\quad W_{ab}=\tr[\rho \left(
J_{a}J_{b}+J_{b}J_{a}\right)] -\delta _{ab}.
\end{equation}
According to \cite{Giraud08}, $\rho$ is classical if and only if
the real symmetric $3\times 3$ matrix $Z$ with matrix elements
\begin{equation}
\label{condC}
Z_{ab}=W_{ab}-u_a u_b
\end{equation}
is non-negative.

Let $\ket{\psi}$ be a fixed pure spin-1 state. Its Majorana representation
consists of two points on the sphere. Since any state obtained by a global
rotation of these two points is at the same distance from $\calc$, one can
without loss of generality consider that these two points are specified by
angles $(\ovtheta,0)$ and $(\pi-\ovtheta,0)$. In the $\ket{jm}$ basis, the
corresponding state is given by (\ref{canonicalj1}),
with $\sin\ovtheta=\sqrt{2}/x$, since the
Majorana polynomial of state $\ket{\psi_x}$ is given by
$M(\zeta)=(1+\sqrt{2}x\zeta+\zeta^2)/\sqrt{x^2+2}$. When
$\ovtheta$ 
varies from $0$ to $\pi/2$, $x$ takes values between $\sqrt{2}$ and
$\infty$. Any spin-1 state can thus be brought to the canonical form
\eref{canonicalj1} with $x\in [\sqrt{2},\infty[$.\\

{\bf Lemma.} The state defined by two antipodal points on the Majorana sphere
is at distance $\sqrt{3/8}$ from $\calc$.\\

\noindent
{\it Proof.} Without loss of generality we take the two points to be at the
North and South poles. The corresponding state is $\ket{1,0}$. Its
parameters in the expansion \eref{canonrhoj1} are $\bfu=0$ and
$W=$diag$(1,1,-1)$.
A coherent state has parameters $\bfu=\bfn$ and $W_{ab}=n_an_b$, where
$\bfn$ is a three-dimensional unit vector which we parametrize as
\begin{equation} \label{nvec}
\bfn=(\sin\theta\cos\varphi,\sin\theta\sin\varphi,\cos\theta)^T\,.
\end{equation}
Since $\calc$ is the convex
hull of the set of coherent states, any element $\rho_c(\bfu,W)$ of
$\calc$ can be written as
\begin{eqnarray}
\label{equations_spin1}
\bfu&=&\sum_{i=1}^N\lambda _{i}\bfni,\\
W_{ab}&=&\sum_{i=1}^N\lambda_i(n_i)_a(n_i)_b,
\end{eqnarray}
where $(n_i)_a$, $a=x,y,z$, are the components  of the vector $\bfni$ and
$N$ is an integer.
The distance $||\rho-\rho_c||^2$ with $\rho=\ket{1,0}\bra{1,0}$ is
\begin{equation}
\label{d2a}
\frac{1}{2}\sum_{i,j}\lambda_i\lambda_j\bfni.\bfnj+
\frac{1}{4}\sum_{i,j}\lambda_i\lambda_j|\bfni.\bfnj|^2
-\frac{1}{2}\sum_i\lambda_i\sum_{a,b}W_{ab}(n_i)_a(n_i)_b+\frac{3}{4}\,.
\end{equation}
Using $\sum_i\lambda_i=1$ and symmetrizing the last term but one
in \eref{d2a} we get
\begin{eqnarray}
\label{d2b}
||\rho-\rho_c||^2&=&\frac{1}{4}\sum_{i,j}\lambda_i\lambda_j(2\bfni.\bfnj+
|\bfni.\bfnj|^2\\
&-&\sum_{a,b}W_{ab}(n_i)_a(n_i)_b-\sum_{a,b}W_{ab}(n_j)_a(n_j)_b+3).\nonumber
\end{eqnarray}
We parametrize vectors $\bfni$ by angles $\theta_i$ and $\varphi_i$ as in (\ref{nvec}). After some
trigonometric simplifications we obtain 
\begin{eqnarray}
&||\rho-\rho_c||^2&=\frac{3}{8}+\frac{1}{4}\sum_{i,j}\lambda_i\lambda_j\left(\frac{1}{2}
\sin^2\theta_i\sin^2\theta_j\cos2(\varphi_i-\varphi_j)\right.\nonumber\\
&+&2\sin\theta_i\sin\theta_j\cos(\varphi_i-\varphi_j)
(1+\cos\theta_i\cos\theta_j)\\ 
&+&\left.
2\cos\theta_i\cos\theta_j+\frac{3}{4}(2-\sin\theta_i-\sin\theta_j)
+\frac{3}{2}\cos^2\theta_i\cos^2\theta_j\right).\nonumber\label{d2c}
\end{eqnarray}
All terms in \eref{d2c} are positive or can be
expanded as a sum of terms of the form $\left(\sum_i\lambda_i
f(\theta_i)\right)^2$. 
Thus $||\rho-\rho_c||^2\geq 3/8$ for all states $\rho_c$. 
This minimum is reached if all terms in \eref{d2c}
vanish, which implies the conditions $\theta_i=\pi/2$ and
\begin{eqnarray}
\sum_i\lambda_i\cos\varphi_i=0,&\ \ \ \  \sum_i\lambda_i\cos 2\varphi_i=0\\
\sum_i\lambda_i\sin\varphi_i=0,&\ \ \ \  \sum_i\lambda_i\sin 2\varphi_i=0.
\end{eqnarray}
These equations admit the solution $\lambda_i=1/N$ and $\varphi_i=2\pi i/N$
for $N\geq 3$, which corresponds to state
\begin{equation}
\rho_c=\left(\begin{array}{ccc}\frac{1}{4}&&0\\
&\frac{1}{2}&\\
0&&\frac{1}{4}\end{array}\right).\label{rhocj1}
\end{equation}
Since $\rho_c\in\calc$, the minimum $3/8$ is indeed reached. Thus
the distance between $\ket{1,0}\bra{1,0}$ and $\calc$ is $\sqrt{3/8}$.
\hfill$\Box$\\

{\bf Theorem.} The state $\ket{1,0}$ is the unique pure QQ state up to
rotations. Its Majorana 
representation is given by a pair of antipodal points.\\

\noindent
{\it Proof.} 
Since $\ket{1,0}$ is at distance $\sqrt{3/8}$ from $\calc$, it
suffices to show that any other pure state is at distance smaller than
$\sqrt{3/8}$, for instance by explicitly exhibiting a classical state which is closer.
We distinguish two cases. For $x\geq \sqrt{6}$ one can show,
using the $Z$-criterion \eref{condC}, that the state
\begin{equation}
\label{rhop}
\rho_c(x)=\left(\begin{array}{ccc}
\frac{1}{4}&a(x)&b(x)\\
a(x)&\frac{1}{2}&a(x)\\
b(x)&a(x)&\frac{1}{4}\end{array}\right)
\end{equation}
with $a(x)=x/(x^2+2)$ and $b(x)=1/(x^2+2)$ is classical. The distance
\begin{equation}
||\ket{\psi_x}\bra{\psi_x}-\rho_c(x)||^2=\frac{3}{8}\left(\frac{x^2-2}{x^2+2}\right)^2\label{distancex}
\end{equation}
is a strictly increasing function of $x$ on $[\sqrt{6},\infty[$, and thus
any state $\ket{\psi_x}$ with $\sqrt{6}\leq x <\infty$ is at distance
squared less than $3/8$.

For $x\leq\sqrt{6}$, one can similarly 
show that the state given by \eref{rhop}
with $a(x)=x/(x^2+2)$ and $b(x)=4(x/(x^2+2))^2-1/4$ is classical, and that its
squared distance to $\ket{\psi_x}$,
\begin{equation}
||\ket{\psi_x}\bra{\psi_x}-\rho_c(x)||^2=\frac{(x^2-2)^2(x^4+12)}{2(x^2+2)^4},
\end{equation}
is a strictly increasing function of $x$ on $[\sqrt{2},\sqrt{6}]$, thus
bounded by 
its value at $\sqrt{6}$, which is $3/32$. 
Thus, the state with $x=\infty$ and the states obtained from it by rotations
are the only pure states at distance $\sqrt{3/8}$, all 
other being closer, which completes the proof.
\hfill$\Box$\\

Numerical evidence shows
that the distance 
given by (\ref{distancex}) is precisely the distance between $\ket{\psi}$
and $\cC$ for all $x$ (see figure \ref{figure0}).

\subsubsection{Mixed states}
The above Theorem shows that only states of the form  $R\ket{1,0}$, where
$R$ is a rotation of the coordinate  frame, are pure QQ states for spin
1. Since pure QQ states are at a distance $Q^2=3/8$ from $\calc$, any mixed
QQ state has to be at least at the same distance.  But, as explained in
point 4.~of section \ref{sec.defprop}, convexity implies that mixed states
can never be further away from $\calc$ than the pure states they are composed
of. Therefore, only mixtures $\rho$ of pure
QQ states that verify $Q(\rho)^2=3/8$ are candidates for mixed 
QQ states. 

Let $\rho$ be a mixed QQ state. Its most general form of $\rho$ is
\begin{equation}
\label{expandrho}
\rho=\sum_i\mu_iR_i\ket{1,0}\bra{1,0}R_i^{\dagger},
\end{equation}
where $R_i$ represents an arbitrary rotation of the coordinate frame, and
the $\mu_i$ are positive and sum up to 1. The state
\begin{equation}
\rho_c=\sum_i\mu_iR_i\left(\begin{array}{ccc}\frac{1}{4}&0&0\\0&\frac{1}{2}&0\\0&0&\frac{1}{4}\end{array}\right)R_i^{\dagger}
\end{equation} 
belongs to $\calc$ and, by convexity of the norm, satisfies
$||\rho-\rho_c||\le\sqrt{3/8}$.  Since $Q(\rho)=\sqrt{3/8}$, $\rho_c$ is
indeed the classical state closest to $\rho$, which implies that $\rho_c$
has to be on the 
boundary of $\calc$. From \eref{uW},
coordinates $\bfu$ and $W$ are linear in $\rho$. Moreover, they transform
respectively as a vector and a 2-tensor under rotations (see
e.g.~\cite{Ritter05}). Using the fact that coordinates for state
diag($1/4,1/2,1/4)$ read $\bfu={\bf 0}$ and $W=$diag($1/2,1/2,0)$, for
$\rho_c$ we have $\bfu(\rho_c)={\bf 0}$ and
\begin{equation}
\label{Wrho}
W(\rho_c)=\sum_i\mu_i
R_i\left(\begin{array}{ccc}\frac{1}{2}&0&0\\0&\frac{1}{2}&0\\0&0&0\end{array}\right)R_i^{\dagger}.
\end{equation}
The set $\calc$ is the set of density matrices such that $Z\geq 0$.
Since $\rho_c$  is on the boundary of $\calc$ its matrix $Z$, which is equal
to $W(\rho_c)$, has a vanishing eigenvalue. Thus there exists some vector
$\bfn$ 
such that $\sum_{ab}W_{ab}n_an_b=0$. Using \eref{Wrho} one easily concludes
that for all rotations, $R_i\bfn$ is equal to $(0,0,1)^T$. Thus either all
rotations are equal or they have the same rotation axis $(0,0,1)^T$. In the
latter case the state $\ket{1,0}$ is invariant under $R_i$, and in both
cases one concludes from \eref{expandrho} that $\rho$ is a pure state.

Thus  $\ket{1,0}$ is the unique spin-1 QQ state up to rotation.

\subsection{Higher values of $j$}
\label{higherj}
Again we concentrate on pure states.
The problem reduces to finding the maximum over all pure states $\ket{\psi}$ of
the minimum over $\rho_c\in\calc$ of
\begin{equation}
\label{tracegen}
\tr\left(\ket{\psi}\bra{\psi}-\rho_c\right)^2=
1-2\sum_i\lambda_i|\langle\psi|\alpha_i\rangle|^2+
\sum_{i,k}\lambda_i\lambda_k|\langle\alpha_i|\alpha_k\rangle|^2\,,
\end{equation}
where $\ket{\alpha_i}=\ket{\ovtheta_i\varphi_i}$ are coherent states. 
The last term in \eref{tracegen} involves the overlap between
coherent states
\begin{equation}
|\langle\alpha|\alpha'\rangle|^2=\cos^{4j}
\frac{\gamma(\alpha,\alpha')}{2},
\end{equation}
where 
$\gamma(\alpha,\alpha')$ is the angle between the two points
corresponding to $\bra{\alpha}\bJ\ket{\alpha}$ and
$\bra{\alpha'}\bJ\ket{\alpha'}$. 
The other
sum in 
\eref{tracegen} involves the overlap between $\ket{\psi}$ and the
coherent states
\begin{equation}
|\langle\psi|\alpha\rangle|^2=
\frac{|\psi_j|^2}{\prod_{i=1}^{2j}\cos^2\frac{\ovtheta_i}{2}}
\prod_{i=1}^{2j}\sin^2\frac{\gamma(\alpha,\zeta_i)}{2},
\end{equation}
where $\zeta_i$ are the Majorana points corresponding to state
$\ket{\psi}$ (for simplicity of notation, we assume that $\psi_j\ne 0$)
and $\gamma(\alpha,\zeta_i)$ is the angle between the point $\bra{\alpha}\bJ\ket{\alpha}$
and the Majorana point $\zeta_i$. Since $\zeta_i$ are the roots of the Majorana 
polynomial, whose coefficients depend on the components $\psi_i$ of
$\ket{\psi}$, it is possible to show, using coefficient-root relations
and normalization of $\ket{\psi}$, that
\begin{equation}
\frac{|\psi_j|^2}{\prod_{i=1}^{2j}\cos^2\frac{\ovtheta_i}{2}}=
\frac{\prod_{i=1}^{2j}(1+|\zeta_i|^2)}{\sum_{k=0}^{2j}|\sigma_k|^2/{2j
    \choose k}}
\end{equation}
with $\sigma_k$ the $k$th symmetric polynomial of the $\zeta_i$ ($\sigma_0=1$,
$\sigma_1=\sum_i\zeta_i$, $\sigma_2=\sum_{i<j}\zeta_i \zeta_j$,...).
For the lowest values of $j$ one can express this quantity as a function of
the angles $\gamma_{ik}$ between points $\zeta_i$ and $\zeta_k$. It is equal
to $1$ for $j=1/2$. For $j=1, 3/2$, we have
\begin{equation}
\frac{|\psi_j|^2}{\prod_{i=1}^{2j}\cos^2\frac{\ovtheta_i}{2}}=1-\frac{1}{2j}\sum_{1\leq i<k\leq 2j}\sin^2\frac{\gamma_{ik}}{2}\,.
\end{equation}
For $j=2, 5/2$, 
\begin{eqnarray}
\frac{|\psi_j|^2}{\prod_{i=1}^{2j}\cos^2\frac{\ovtheta_i}{2}}&=&1-\frac{1}{2j}\sum_{1\leq i<k\leq 2j}\sin^2\frac{\gamma_{ik}}{2} \\
&+&\frac{1}{2j(2j-1)}\sum_{\textrm{all pairwise}}\sin^2\frac{\gamma_{ik}}{2}
\sin^2\frac{\gamma_{i'k'}}{2},\nonumber
\end{eqnarray}
with the last sum running over all ways of
pairing $2j$ points into two distinct pairs. These formulae should easily
generalize 
to general $j$. The whole expression \eref{tracegen} can thus be
expressed as a function of terms of the form $\sin(\gamma/2)$, which are equal to half
the Euclidean distance between a pair of points separated by an angular
distance $\gamma$. For instance in the case $j=1$ the problem corresponds to
finding
\begin{equation}
\max_{\zeta_i}\min_{\lambda_i,\alpha_i}\left(1
-2\sum_i\lambda_i\frac{\sin^2\frac{\gamma(\alpha_i,\zeta_1)}{2}\sin^2\frac{\gamma(\alpha_i,\zeta_2)}{2}}{1-\frac{1}{2}\sin^2
\frac{\gamma(\zeta_1,\zeta_2)}{2}}
+\sum_{i,k}\lambda_i\lambda_k\cos^{4j}\frac{\gamma(\alpha_i,\alpha_k)}{2}\right)\,.\label{minmax}
\end{equation}

Our quest for pure QQ states thus amounts 
to finding an optimal arrangement of points on the sphere with
two types of particles $\zeta_i$ and $\alpha_i$. 
The problem of arranging points on a sphere as evenly as possible has a long
history. It was known by the ancient Greeks that 4,6,8,12 or 20 points could
be arranged in a regular way. About 25 centuries later,
by classifying all finite subgroups of the group of rotations in
$\mathbb{R}^3$, it was proved that only five regular polyhedra exist.
In the framework of electrostatics, one can define a generalized
Coulomb potential between $n$ point charges on the sphere as
\begin{equation}
\label{coulomb}
\sum_{i\neq j}\frac{1}{d_{ij}^m}
\end{equation}
where $d_{ij}$ is the distance between points $i$ and $j$, and $m$ a positive
integer. The question of finding a configuration of points on the sphere
that minimizes the potential \eref{coulomb} was first investigated by
Thomson \cite{Thomson1904}. Similar questions appear in many fields, from crystallography to
biology (see \cite{Altschuler97} and references therein).
Our problem bears some similarity with such questions. However, our
potential is more complicated (see e.g.~\eref{minmax}) and two
kinds of ``particles'' are involved. Intuitively, for fixed $\zeta_i$,
the minimization problem in \eref{minmax} corresponds to finding
coherent states $\alpha_i$ as remote as possible from the $\zeta_i$
and from each other.

Given the complexity of the minimax problem of the kind of
(\ref{minmax}) beyond the case $j=1$, we choose a numerical
approach. Many algorithms were devised to numerically obtain optimal configurations of
points. Rather surprisingly it turns out that the optimal distribution does
not necessarily coincide with regular polyhedra even in the case where these
exist (see e.g.~\cite{Edmundson92}, where the distribution of point
charges that minimizes Coulomb potential \eref{coulomb} is given up to
$60$ points). In the next subsections we will apply numerical
techniques to identify QQ states for the smallest Hilbert dimensions.

\subsection{Numerical procedure}\label{numerics}
The problem of finding the QQ states can be reformulated in terms of convex
optimization, and even as an instance of quadratic programming.

\subsubsection{Quadratic programming}
For a fixed state $\rho\in\caln$, we represent the matrix $\rho
_{c}\in\calc$ minimizing the distance to $\calc$ as a linear combination of
coherent  
states whose directions densely and uniformly cover the unit sphere, 
$\rho _{c}=\sum_{i=1}^{N}\lambda _{i}\left\vert \alpha_{i}\right\rangle
\left\langle \alpha_{i}\right\vert $ with $N$ large.  $Q^{2}\left( \rho
\right) $ then becomes a quadratic function of the coefficients $\lambda
_{i}$ which has to be minimized under the constraints $\lambda
_{i}\geq 0,\sum_{i=1}^{N}\lambda _{i}=1$. This is a problem of
quadratic programming which can be solved by a variety of
algorithms. Although the original linear combination contains several
thousand coherent states, only a few of them enter the solution
with coefficients significantly different from zero. In the search
of QQ states, the result of the quadratic minimization was then
numerically maximized by variation of the pure state
$\rho$. The Majorana configurations
were found numerically and then deformed to closest symmetrical figures
under the condition that $Q$ increased. It    
is probably superfluous to stress that maximization and
minimization do not commute, so that the maximin and minimax of
the squared distance $\left\Vert \rho -\rho _{c}\right\Vert^2$ are
different.

The optimization algorithm itself starts from $N$ coherent states
randomly distributed on a sphere, on which quadratic programming
is performed. This yields an intermediate optimal state which is a
combination of a relatively small number $M\simeq 5-25$, $M\ll N$,
of coherent states. If some of these coherent states point in
directions closer than a certain threshold, say, 2 degrees, they
are replaced by a single coherent state with a cumulative weight
directed along the weighted average direction. This step yields
$M'$ coherent states. Then $N-M'$ new random coherent states are
generated and the  quadratic programming algorithm is run again
starting from these $N$ coherent states ($N-M'$ new and $M'$ old).
Iterating the process $K$ times (typically $K~1000-5000$) with
$N~100$ yielded a 7-8 digit precision of the squared distance
value.

As expected from the remark of section \ref{rhocisev}, the numerically found
wavefunction  of a QQ state 
for given $j$ coincides, within numerical accuracy, with an eigenvector of
the density matrix 
 of its nearest classical neighbours. In fact, the
accuracy with which this property was fulfilled could serve as a
measure of accuracy in the search of the maximin. 
An interesting point is that all QQ states that we obtained were invariably 
associated with the eigenvector of the density matrix of the
classical state corresponding to its largest eigenvalue.

\bigskip

\subsection{Results}
\begin{figure}
\begin{center}
\includegraphics[scale=1.]{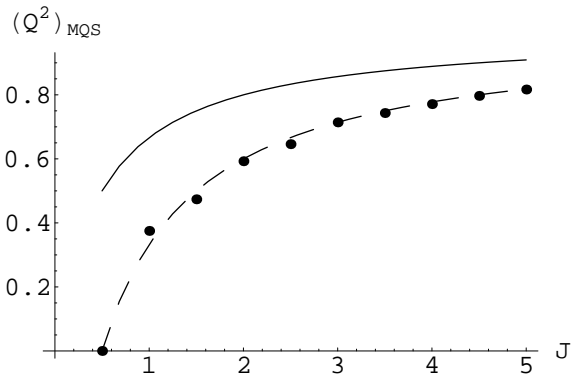}
\end{center}
\caption{Squared distance of the Queen of Quantum states from the
boundary of the classical domain $\calc$ for $2j=1-10$ (dots), the empirical
approximation $1-2/(2j+1)$ (dashed line) and the rigorous upper bound 
$1-1/(2j+1)$(full line).\label{distance}}
\end{figure}

\begin{figure}
\begin{center}
  \includegraphics[scale=0.7]{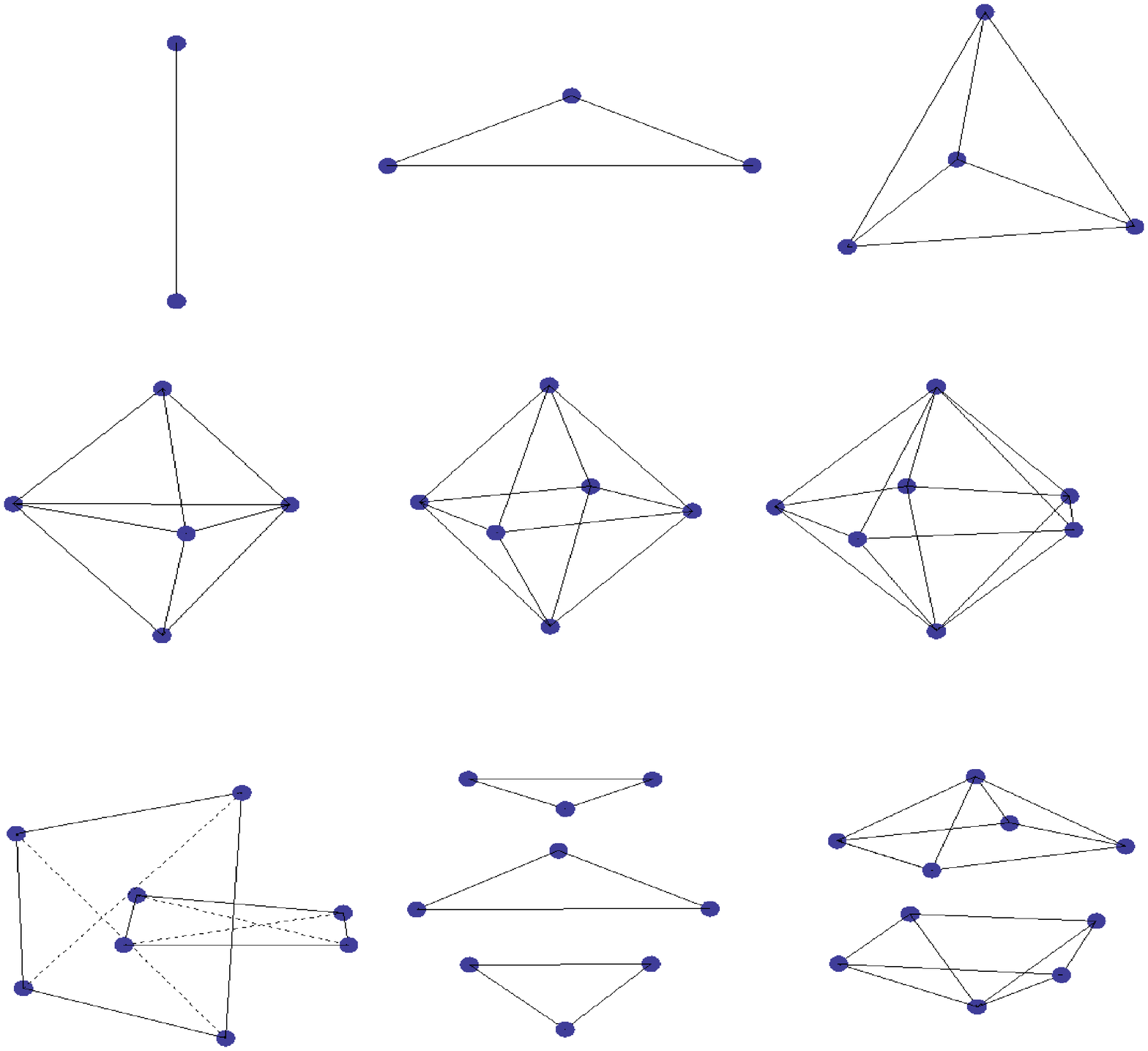}
\end{center}
\caption{Queens of Quantum: The Majorana points for $j=1, 3/2, 2$ (top), 
$j=5/2, 3, 7/2$ (middle), $j=4, 9/2, 5$ (bottom). See Table \ref{majorana_app}
  for a precise definition.\label{majoplots}}
\end{figure}

\begin{figure}
\begin{center}
  \includegraphics[scale=0.7]{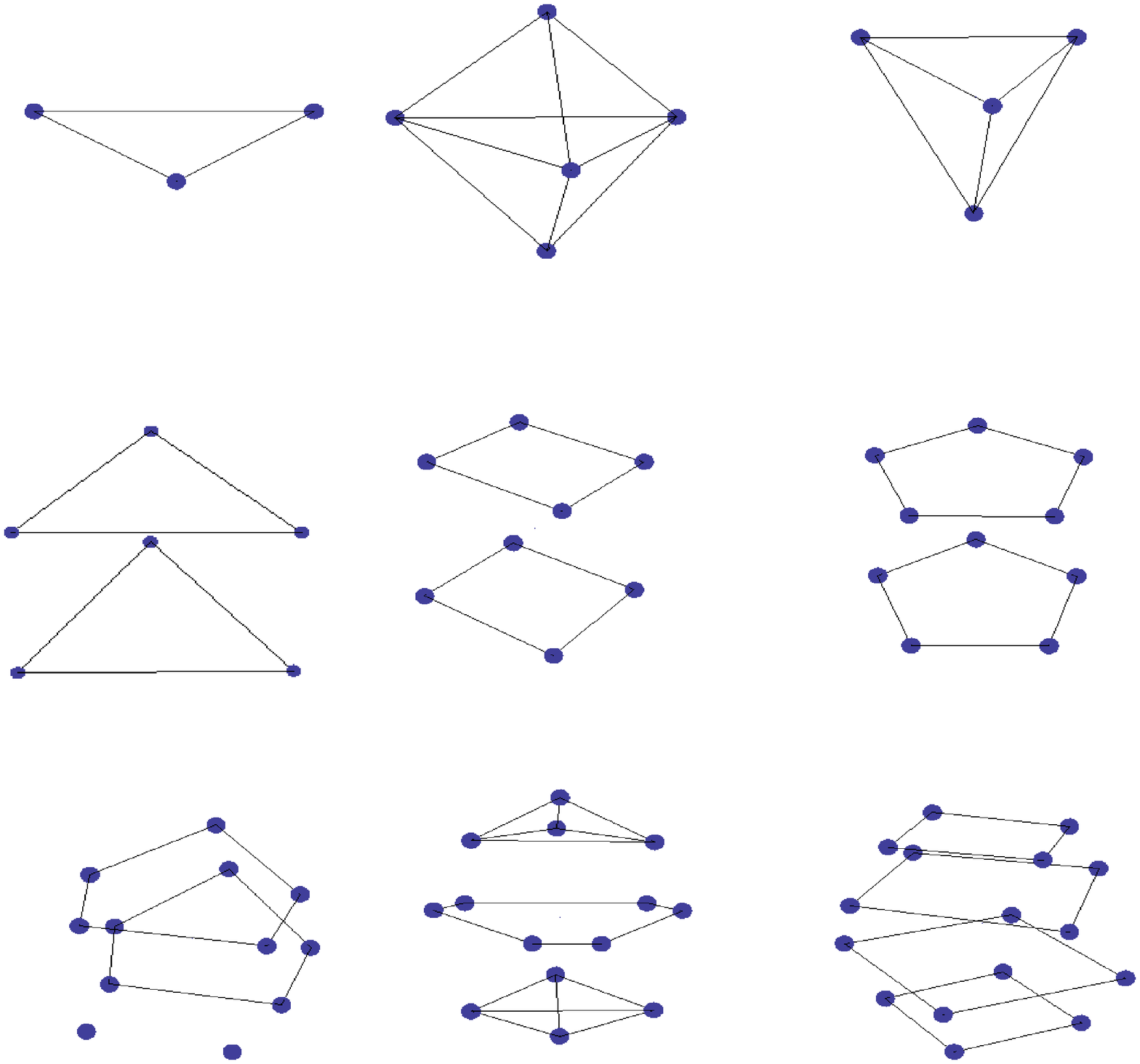}
\end{center}
\caption{Classical states closest to the Queens of Quantum: the 
coherent state points $(\ovtheta, \varphi)$
for $j=1, 3/2, 2$ (top), $j=5/2, 3, 7/2$ (middle), $j=4, 9/2, 5$ (bottom). 
See Tables \ref{coherent_app} and \ref{coherent_app2} 
for a precise definition.\label{cohplots}}
\end{figure}

We carried out a numerical search of the QQ states for 
$j$ from $1/2$ to $5$. The resulting values of the distance 
are plotted in figure \ref{distance} together with the upper
bound (\ref{upper_bound_pure}). 
The distance of the QQ state from
$\mathcal{C}$ grows monotonically with the increase of $j$; an
almost perfect fit is
\begin{equation}
   Q_{QQ}^2\approx 1-2/(2j+1).
\end{equation}
 The resulting
arrangements of the Majorana points on the unit sphere that represent the QQ
states, as well as the respective sets of coherent states constituting the
nearest $\rho _{c}$ are given in Table~\ref{majorana}.
The numerically obtained values for the
Majorana points and coherent states are listed in the Appendix (Tables
\ref{majorana_app} and \ref{coherent_app},\ref{coherent_app2}). For the  
first few values of $j$ it is possible to identify regular structures from
these numerical results. They can be recognized as highly symmetric
figures (see figures \ref{majoplots} and \ref{cohplots}) in which $\ovtheta$ and $\varphi$ are
typically rational multiples of $\pi$. 
Table~\ref{majorana} gives the
analytical 
expression that we obtain for the QQ states in the $\ket{jm}$ basis 
if we identify the numerical results with these regular structures.  
We find in particular that for $j=2$ and $j=3$ (i.e.~4 and 6 Majorana
points), the QQ states correspond to the Platonic bodies (tetrahedron and
octahedron, respectively). For $j=4$, where one would expect to
find the next Platonic body, the cube, the symmetry of the Majorana
configuration is lower (see figure \ref{majoplots}).

The number of local
maxima of $Q^{2}$ with close values  rapidly 
increases with the growth of $j$ while the maxima themselves tend
to become more and more shallow. This makes the search of the
optimal configuration more and more difficult while the value of
the maximin itself can still be reliably established.\\

It is instructive to compare our results with the optimal
distribution of identical point charges on a unit sphere which interact
through the standard Coulomb repulsion, plotted e.g. 
in~\cite{Thomson1904}.
The analogy between the problems follows from the possibility to
express the optimized quantity in terms of distances between the
end points of the Majorana vectors and the vectors of the coherent
states (see section~\ref{higherj}). 
In the range $j=1-5$ the symmetry of the 
Majorana configurations  of the QQ states coincides with that of the
equilibrium configurations of $2j$ charges on a sphere, the only
exception being $j=4$. Indeed, the optimal Coulomb
configuration of 8 identical point charges is the square antiprism with symmetry $D_{4d}$.
However that latter configuration gives only a local maximum of
$Q^{2}$ equal to $0.76868$  which is slightly less than the global
maximum $0.77108$  realized in a configuration with  lower
symmetry $D_{2d}$ (Table \ref{majorana}).\\

It is also interesting to compare our results to the anticoherent spin
states  introduced in \cite{Zimba06}. These states are defined
such that 
$\langle {\bf J}\rangle=0$ and 
$(\Delta{\bf J_n})^2=\langle {\bf n.J}{\bf n.J}\rangle-\langle {\bf
  n.J}\rangle\langle{\bf n.J}\rangle$ is uniform over the unit sphere,
i.e. independent of ${\bf n}$. Platonic states for $j=2,3,4,6$ and $10$ are
shown to be anticoherent.  In \cite{Kolenderski08} it was shown that
multiqubit states with diagonal spin covariance matrix  
and maximal variances of each spin component are optimal for
reference frame alignment. This property is verified for our QQ states for
$j=2,3$. QQ states with $j=2,3$ are therefore both 
``anticoherent'' and optimal for reference frame alignment. 

\begin{center}
\begin{table}
\begin{tabular}{|c|c|p{5cm}|c|p{4.5cm}|}
\hline $j$&$Q^2_{\mathrm{QQ}}$& Majorana points of $\ket{\psi}$ on unit
sphere 
&QQ state&Coherent state points of closest classical neighbours\\
\hline \hline $1$&$3/8$ &Two antipodal
points&$\frac{1}{\sqrt{2}}\left(\ket{1,1}+\ket{1,-1}\right)$&Equilateral
triangle in
equatorial plane\\
\hline $3/2$&$9/19$ &Equilateral triangle in equatorial plane
&$\frac{1}{\sqrt{2}}\left(\ket{\frac{3}{2},\frac{3}{2}}+\ket{\frac{3}{2},-\frac{3}{2}}\right)$&Two
points on poles,
equilateral triangle in equatorial plane\\
\hline $2$& $16/27$ &Tetrahedron&
$\sqrt{\frac{2}{3}}\ket{2,1}+\sqrt{\frac{1}{3}}\ket{2,-2}$&Overturned tetrahedron\\
\hline $5/2$&0.645914$^a$&Two points on poles, equilateral triangle in
equatorial
plane&$\frac{1}{\sqrt{2}}\left(\ket{\frac{5}{2},\frac{3}{2}}+\ket{\frac{5}{2},-\frac{3}{2}}\right)$&Two
parallel equilateral triangles symmetric on both sides 
of  equatorial plane\\
 \hline $3$&$347/486$ &
Octahedron&$\frac{1}{\sqrt{2}}\left(\ket{3,2}+\ket{3,-2}\right)$&Cube\\
 \hline $7/2$ &0.743138$^b$& Two points on poles, regular pentagon in
equatorial plane
&$\frac{1}{\sqrt{2}}\left(\ket{\frac{7}{2},\frac{5}{2}}+\ket{\frac{7}{2},-\frac{5}{2}}\right)$
 &Two
 parallel regular pentagons\\
 \hline $4$ &0.77108
 & Four points in  plane with line of symmetry; remaining four points obtained by 
improper
 $\pi/2$-rotation $S_4$ about symmetry line&see Table~\ref{majorana_app} &Twelve coherent states\\
\hline $9/2$ &0.79676 &Three triangles in parallel planes,
central  rotated by $\pi$
&see Table~\ref{majorana_app}&Two points on poles, two triangles symmetric with respect to equatorial plane
and  three doublets on equator\\
\hline $5$ &0.81664 & Two points on poles and two squares in
parallel
 planes rotated with respect to each other by $\pi/4$
 &see Table~\ref{majorana_app}&Four squares in parallel planes;
the two   South squares rotated by $\pi/4$ with respect to those in North
hemisphere\\
\hline
\end{tabular}
\caption{Queens of Quantum: The Majorana points of the QQ states, the QQ
  states in $|jm\rangle$ notation, and the set of
coherent states of the nearest classical neighbour.\\
$^a$ Minimum of $(270286+61910\cos(2x)+58680\cos(4x) +
855\cos(6x)+1530\cos(8x)-45\cos(10x)-51200\sin(x)+25600\sin(3x)-5120\sin(5x))/262144$\\
$^b$ Minimum of
$(68477212 + 10990343\cos(2x) + 18268726\cos(4x) + 2030189\cos(6x) + 
    845124\cos(8x) + 25319\cos(10x) + 26474\cos(12x) - 91\cos(14x) - 
    4014080\sin(x) + 2408448\sin(3x) - 802816\sin(5x) + 114688\sin(7x))/67108864$
}
\label{majorana}
\end{table}
\end{center}

\section{Conclusion}
In summary, we have introduced the ``quantumness'' $Q(\rho)$ for any
finite-dimensional quantum state $\rho$. This quantity measures how
quantum a state is. $Q(\rho)$ is a real valued,
positive, and  convex function with a value
between 0 and 1 (more precisely, $0\le Q(\rho)^2<1-1/d$, where $d$ is the 
dimension of the Hilbert space) that measures the Hilbert-Schmidt distance
of $\rho$ to the convex set of classical states, defined as 
states with positive $P$-function \cite{Giraud08}. We have shown that
thermal states always become classical ($Q(\rho)=0$) for temperatures larger
than a critical temperature that depends on the dimension of the Hilbert
space and the Hamiltonian, whereas the ground state of a system may or may
not have non-zero  
quantumness. We used $Q(\rho)$ in order to find the ``Queens of Quantum''
states, defined as the states with maximum quantumness for a given Hilbert
space dimension. Maximum quantumness can always be reached for pure states,
and we have demonstrated that the Queens of Quantum states correspond to
beautiful, highly symmetric bodies when expressed in terms of their Majorana
representation. For the two lowest dimensions that allow for Platonic bodies
($j=2$ and 3, with 4 and 6 Majorana points, respectively), they are indeed
the corresponding Platonic bodies (tetrahedron and octahedron,
respectively). For $j=4$, lowering the symmetry allows to obtain an even
larger quantumness compared to the one for the corresponding Platonic body
(the cube), and we have identified numerically all other Queens of Quantum
states for $3/2\le j\le 5$ using quadratic programming.

\ack
This work was supported in part by the Agence Nationale de la Recherche
(ANR), project QPPRJCCQ. PB is grateful to the Sonderforschungsbereich TR 12 
of the Deutsche Forschungsgemeinschaft and to the GDRI-471.

\appendix
\setcounter{section}{1}
\section*{Appendix: QQ states and their closest classical state for $j\leq 5$}
This appendix lists the numerical results obtained from the algorithms
explained in section \ref{numerics}. 
The Majorana configurations were found numerically and
then deformed to the closest symmetrical figures under the condition that
$Q$ was 
increased. Results are shown in Table~\ref{majorana_app}.
The coherent states shown in Tables~\ref{coherent_app},\ref{coherent_app2}
were numerically found 
for these symmetrical configurations.  Contrary to the Majorana points, the
symmetry of the coherent states was not enforced.  Along with the numerically
found values, we also give the likely exact values for the coherent states
and the weights in Tables~\ref{coherent_app},\ref{coherent_app2}, as far as
they can be deduced from the numerically found ones and symmetry
considerations. 
\begin{center}
\begin{table}
\begin{tabular}{|c|c|c||c|c|c|}
\hline 
$j$&$\ovtheta$&$\varphi$& $j$&$\ovtheta$&$\varphi$\\
\hline \hline 
$1$&0&$\pi$&$4$& $\pi$/2& 4.46095\\
&$\pi$&$\pi$&& $\pi$/2&1.82223\\
\cline{1-3}
$3/2$& $\pi$/2& $\pi$&& $\pi$/2& 5.62398\\
& $\pi$/2&5$\pi$/3 && $\pi$/2&0.659206\\
& $\pi$/2& $\pi$/3&& 0.251438& 0\\
\cline{1-3}
$2$& 0&$\pi$ && 2.890154& 0\\
& 2$\arccos(1/\sqrt{3})$& $\pi$&& 0.911591& $\pi$\\
& 2$\arccos(1/\sqrt{3})$& 5$\pi$/3&& 2.230002& $\pi$\\
\cline{4-6}
& 2$\arccos(1/\sqrt{3})$&4$\pi$/3 &$9/2$& 0.799772& $\pi$\\
\cline{1-3}
$5/2$& 0&$\pi$ && 0.799772& 5$\pi$/3\\
& $\pi$/2& $\pi$&& 0.799772& $\pi$/3\\
& $\pi$/2& 5$\pi$/3&& $\pi$/2& 0\\
& $\pi$/2& $\pi$/3 && $\pi$/2& 2.0944\\
& $\pi$3& $\pi$&& $\pi$/2& 4$\pi$/3\\
\cline{1-3}
$3$& 0&$\pi$ && 2.341821& $\pi$\\
& $\pi$/2& $\pi$&& 2.341821& 5$\pi$/3\\
& $\pi$/2& 4$\pi$/3&& 2.341821& $\pi$/3\\
\cline{4-6}
& $\pi$/2& 0&$5$& 0&  $\pi$\\
& $\pi$/2& 4$\pi$/3&& 1.134586& $\pi$\\
& $\pi$3& $\pi$&& 1.134586& 3$\pi$/2\\
\cline{1-3}
$7/2$& 0& $\pi$&& 1.134586& 0\\
& $\pi$/2& $\pi$&& 1.134586& $\pi$/2\\
& $\pi$/2& 7$\pi$/5&& 2.007007& 5$\pi$/4\\
& $\pi$/2& 9$\pi$/5&& 2.007007& 7$\pi$/4\\
& $\pi$/2& $\pi$/58&& 2.007007& $\pi$/4\\
& $\pi$/2& 3$\pi$/5&& 2.007007& 3$\pi$/4\\
& $\pi$& $\pi$&& $\pi$& $\pi$\\
\hline
\end{tabular}
\caption{Numerical coordinates $(\ovtheta,\varphi)$ of the Majorana points for
the Queens of Quantum for $1/2\leq j\leq 5$. The Queens of Quantum are given
by $(\ovtheta,\varphi)$ through \eref{zeta},\eref{Mz}.}
\label{majorana_app}
\end{table}
\end{center}
\begin{table}

\begin{tabular}{|c|c|c|c||c|c|c|}
\hline
$j$&$\ovtheta_i$&$\varphi_i$&$\lambda_i$&$\ovtheta_i^e$&$\varphi_i^e$&$\lambda_i^e$\\
\hline
1& &&
&$\pi/2$&0&1/3\\
& &&&$\pi/2$&$2\pi/3$&1/3\\
& &&&$\pi/2$&$4\pi/3$&1/3\\
\hline
3/2
&0.00194&0.74739&0.39480& 0&irrelevant&\nae\\
&1.56732&6.28061&0.06961& $\pi/2 $&$2\pi$&\nae\\
&1.56840&2.09883&0.07086& $\pi/2$&$2\pi/3$&\nae\\
&1.57863&4.19050&0.07005& $\pi/2$&$4\pi/3$&\nae\\
&3.13684&5.42130&0.39469& $\pi$&irrelevant&\nae\\
\hline
2
&1.23030&0.00030&0.24995& $\pi-2\arccos(1/\sqrt{3})$&0&1/4\\
&1.23069&2.09238&0.25012& $\pi-2\arccos(1/\sqrt{3})$&$2\pi/3$&1/4\\
&1.23256&4.18886&0.24995& $\pi-2\arccos(1/\sqrt{3})$&$4\pi/3$&1/4\\
&3.13661&4.96074&0.24997& $\pi$ &irrelevant&1/4\\
\hline
5/2
&1.11130&2.09564&0.16761& $\arccos(1/\sqrt{5})$&$2\pi/3$&1/6\\
&1.10720&4.19058&0.16749& $\arccos(1/\sqrt{5})$&$4\pi/3$&1/6\\
&1.11397&6.28311&0.16782& $\arccos(1/\sqrt{5})$&0       &1/6\\
&2.03921&2.09192&0.16563& $\pi-\arccos(1/\sqrt{5})$&$2\pi/3$&1/6\\
&2.03993&4.18900&0.16626& $\pi-\arccos(1/\sqrt{5})$&$4\pi/3$&1/6\\
&2.04011&6.28293&0.16519& $\pi-\arccos(1/\sqrt{5})$&$2\pi$  &1/6\\
\hline
3
&0.95491&0.79084&0.12528&$\arccos(1/\sqrt{3})$&$\pi/4$     &1/8\\
&0.95655&2.35809&0.12459&$\arccos(1/\sqrt{3})$&$3\pi/4$    &1/8\\
&0.95934&3.92450&0.12512&$\arccos(1/\sqrt{3})$&$5\pi/4$    &1/8  \\
&0.95581&5.49362&0.12530&$\arccos(1/\sqrt{3})$&$7\pi/4$    &1/8\\
&2.18700&0.77800&0.12498&$\pi-\arccos(1/\sqrt{3})$&$\pi/4$ &1/8\\
&2.18178&2.35353&0.12511&$\pi-\arccos(1/\sqrt{3})$&$3\pi/4$&1/8  \\
&2.19225&3.92883&0.12513&$\pi-\arccos(1/\sqrt{3})$&$5\pi/4$&1/8  \\
&2.18507&5.50033&0.12450&$\pi-\arccos(1/\sqrt{3})$&$7\pi/4$&1/8\\
\hline
7/2
&0.86054&1.25178&0.09995&\nae&$2\pi/5$&1/10\\
&0.85628&2.51288&0.10032&\nae&$4\pi/5$&1/10  \\
&0.85907&3.76843&0.10018&\nae&$6\pi/5$&1/10  \\
&0.86133&5.03012&0.10004&\nae&$8\pi/5$&1/10\\
&0.85692&6.28226&0.09978&\nae&$2\pi  $&1/10\\
&2.28518&1.25083&0.09960&\nae&$2\pi/5$&1/10 \\
&2.28643&2.50451&0.10004&\nae&$4\pi/5$&1/10\\
&2.28247&3.76493&0.10048&\nae&$6\pi/5$&1/10\\
&2.28444&5.03053&0.10013&\nae&$7\pi/5$&1/10\\
&2.28444&6.28132&0.09948&\nae&$2\pi  $&1/10\\
\hline

\end{tabular}

\caption{Numerical coordinates and weights $(\ovtheta_i,\varphi_i, \lambda_i)$ of the
  coherent states for the classical states achieving the minimum distance to
the Queens of Quantum for $1/2\leq j\leq 7/2$. Namely,
$\rho_c=\sum_i\lambda_i\ket{\ovtheta_i\varphi_i}\bra{\ovtheta_i\varphi_i}$
with $\ket{\ovtheta_i\varphi_i}$ given by \eref{zetaC}. The
  $(\ovtheta_i^e,\varphi_i^e, \lambda_i^e)$ are the presumable exact values
  deduced from the numerical data, based on symmetry considerations. In
  various cases their entry reads $\nae$, which means that 
no analytical expression could be extracted from the numerical data. 
In the case $j=1$, the exact analytical solution (\ref{rhocj1}) is given.}
\label{coherent_app}
\end{table}

\begin{table}
\begin{tabular}{|c|c|c|c||c|c|c|}
\hline
$j$&$\ovtheta_i$&$\varphi_i$&$\lambda_i$&$\ovtheta_i^e$&$\varphi_i^e$&$\lambda_i^e$\\
\hline
4
&0.74844&1.95206&0.08164&&&1/12\\
&0.74558&4.34339&0.08225&&&1/12\\
&0.88777&1.24671&0.08324&&&1/12\\
&0.89612&5.02990&0.08139&&&1/12\\
&1.04909&6.27490&0.08630&&&1/12\\
&1.56825&2.62661&0.08525&&&1/12\\
&1.56244&3.66123&0.08541&\nae&\nae&1/12\\
&2.09943&6.28277&0.08604&&&1/12\\
&2.25446&1.25118&0.08245&&&1/12\\
&2.25422&5.03059&0.08205&&&1/12\\
&2.39338&1.94823&0.08187&&&1/12\\
&2.39551&4.33750&0.08211&&&1/12\\
\hline
9/2
&0.00150&1.06854&0.09641&0&irrelevant&\nae\\

&0.83772&4.18728&0.06303&$\arccos(2/3)$&$4\pi/3$&\\
&0.82797&6.27566&0.06297&$\arccos(2/3)$&$2\pi$&\nae\\
&0.82663&2.09864&0.06302&$\arccos(2/3)$&$2\pi/3$&\\

&1.56795&0.83024&0.07166&$\pi/2$&\nae&\\
&1.57137&1.26468&0.07120&$\pi/2$&\nae&\\
&1.57366&2.92169&0.07336&$\pi/2$&\nae&\nae\\
&1.56980&3.37157&0.07006&$\pi/2$&\nae&\\
&1.57408&5.01531&0.07175&$\pi/2$&\nae&\\
&1.57129&5.45752&0.07132&$\pi/2$&\nae&\\

&2.31140&6.28060&0.06284&$\arccos(-2/3)$&$2\pi$&\\
&2.31289&2.08971&0.06326&$\arccos(-2/3)$&$2\pi/3$&\nae \\
&2.30992&4.18140&0.06279&$\arccos(-2/3)$&$4\pi/3$&\\

&3.13728&5.61142&0.09634&$\pi$&irrelevant&\nae\\
\hline
5
&0.73375&0.77724&0.05989 &&$\pi/4$& \\
&0.75268&2.36318&0.05537&\nae&$3\pi/4$& \nae  \\
&0.72910&3.91883&0.05801&&$5\pi/4$& \\
&0.75488&5.50021&0.05713&&$7\pi/4$& \\

&1.27503&0.78214&0.06959&&$\pi/4$&   \\
&1.24043&2.36002&0.07004&\nae&$3\pi/4$& \nae  \\
&1.26250&3.92131&0.07053&&$5\pi/4$& \\
&1.25154&5.48300&0.06869&&$7\pi/4$& \\

&1.88976&6.27189&0.07205&&$2\pi$&  \\
&1.87975&1.57367&0.06859&\nae&$\pi/2$& \nae\\
&1.89339&3.14019&0.07050&&$\pi$& \\
&1.87096&4.70769&0.06717&&$3\pi/2$& \\

&2.42373&6.27224&0.03774&&$2\pi$& \\
&2.38756&1.57823&0.05841&\nae&$\pi/2$& \nae\\
&2.40240&3.14208&0.05641&&$\pi$& \\
&2.38373&4.70632&0.05988&&$3\pi/2$& \\
\hline
\end{tabular}

\caption{Same as Table \ref{coherent_app}, but for $j=4,9/2,5$.}
\label{coherent_app2}
\end{table}

\section*{References}

\bibliography{mybibs_bt}

\begin{thebibliography}{10}

\bibitem{Lewenstein00}
M.~Lewenstein, D.~Bruss, J.~I. Cirac, B.~Kraus, M.~Kus, J.~Samsonowicz,
  A.~Sanpera, and R.~Tarrach.
\newblock {\em J. Mod. Optics}, 47:2841, 2000.

\bibitem{Plenio05}
M.~B. Plenio and S.~Virmani.
\newblock {\em Quantum Information and Computation}, 7:1, April 2005.

\bibitem{Bennett93}
C.~H. Bennett, G.~Brassard, C.~Crepeau, R.~Jozsa, A.~Peres, and W.~K. Wootters.
\newblock {\em Phys. Rev. Lett.}, 70:1895, 1993.

\bibitem{Nielsen00}
M.~A. Nielsen and I.~L. Chuang.
\newblock {\em Quantum Computation and Quantum Information}.
\newblock Cambridge University Press, 2000.

\bibitem{Jozsa03}
R.~Jozsa and N.~Linden.
\newblock {\em Proc. R. Soc. Lond. A}, 459:2011--2032, 2003.

\bibitem{Bell64}
J.~S. Bell.
\newblock {\em Physics}, 1:195, 1964.

\bibitem{Gisin91}
N.~Gisin.
\newblock {\em Phys. Lett. A}, 154(201), 1991.

\bibitem{Kim05}
M.~S. Kim, E.~Park, P.~L. Knight, and H.~Jeong.
\newblock {\em Physical Review A}, 71(4):043805, April 2005.

\bibitem{Mandel86}
L.~Mandel.
\newblock {\em Physica Scripta T}, 12:34, 1986.

\bibitem{Giraud08}
O.~Giraud, P.~Braun, and D.~Braun.
\newblock {\em Phys. Rev. A}, 78(4):042112, 2008.

\bibitem{Davis00}
R.~I.~A. Davis, R.~Delbourgo, and P.~D. Jarvis.
\newblock {\em J. Phys. A: Math. Gen}, 33:1895--1914, 2000.

\bibitem{Zimba06}
J.~Zimba.
\newblock {\em EJTP}, 3:143--156, 2006.

\bibitem{Kolenderski08}
P.~Kolenderski and R.~Demkowicz-Dobrzanski.
\newblock {\em Phys. Rev. A}, 78(5):052333, Nov 2008.

\bibitem{Vedral98}
V.~Vedral and M.~B. Plenio.
\newblock {\em Phys. Rev. A}, 57(3):1619--1633, Mar 1998.

\bibitem{Boyd04}
S.~P. Boyd and L.~Vandenberghe.
\newblock {\em Convex Optimization}.
\newblock Cambridge University Press, 2004.

\bibitem{Agarwal81}
G.~S. Agarwal.
\newblock {\em Phys. Rev. A}, 24:2889, 1981.

\bibitem{Majorana1932}
E.~Majorana.
\newblock {\em Nuovo Cimento}, 9:43--50, 1932.

\bibitem{Penrose89}
R.~Penrose.
\newblock {\em The Emperor's New Mind, Oxford University Press}.
\newblock Oxford, 1989.

\bibitem{Hannay98}
J.~H. Hannay.
\newblock {\em J. Phys. A: Math. Gen}, 31:L53--L59, 1998.

\bibitem{Dennis04}
M.~R. Dennis.
\newblock {\em J. Phys. A: Math. Gen.}, 37:9487, 2004.

\bibitem{Ritter05}
W.~G. Ritter.
\newblock {\em Journal of Mathematical Physics}, 46(8):082103, 2005.

\bibitem{Thomson1904}
J.~J. Thomson.
\newblock {\em Philos. Mag.}, 7:237, 1904.

\bibitem{Altschuler97}
E.~L. Altschuler, T.~J. Williams, E.~R. Ratner, R.~Tipton, R.~Stong, F.~Dowla,
  and F.~Wooten.
\newblock {\em Phys. Rev. Lett.}, 78(14):2681--2685, Apr 1997.

\bibitem{Edmundson92}
J.~R. Edmundson.
\newblock {\em Acta Cryst. A}, 48:60--69, 1992.

\end{thebibliography}

\end{document}